\documentclass[review,12pt]{elsarticle}
\usepackage{lineno,hyperref}
\usepackage{graphicx}
\usepackage{geometry}
\usepackage{caption}
\usepackage{subcaption}
\usepackage[linesnumbered,ruled,lined]{algorithm2e}
\usepackage{float}
\usepackage{tikz}
\usepackage{epstopdf}
\usepackage{amssymb}
\usepackage{amsmath}
\usepackage{indentfirst}
\usepackage{subcaption}
\usepackage{footnote}
\usepackage{array}
\usepackage{cases}
\usepackage{multirow}
\usepackage{ upgreek }
\journal{CMAME}
\begin{document}
%
%
\begin{frontmatter}
		\title{Extended Eulerian SPH and its realization of FVM}
		\author{Zhentong Wang }
		\ead{zhentong.wang@tum.de}
			\author{Chi Zhang}
		\ead{c.zhang@tum.de}
		\author{Oskar J. Haidn}
		\ead{oskar.haidn@tum.de}
		\author{Nikolaus A. Adams}
		\ead{nikolaus.adams@tum.de}
		\author{Xiangyu Hu \corref{mycorrespondingauthor}}
		\ead{xiangyu.hu@tum.de}
		\address{TUM School of Engineering and Design, 
		Technical University of Munich, Garching, 85747, Germany}
		\cortext[mycorrespondingauthor]{Corresponding author. }
\begin{abstract}
Eulerian smoothed particle hydrodynamics (Eulerian SPH) is considered as a potential meshless alternative to a traditional Eulerian mesh-based method, 
i.e. finite volume method (FVM), in computational fluid dynamics (CFD).
 While researchers have analyzed the differences between these two methods, 
 a rigorous comparison of their performance and computational efficiency is hindered 
 by the constraint related to the normal direction of interfaces in pairwise particle interactions within Eulerian SPH framework. 
 To address this constraint and improve numerical accuracy, 
 we introduce Eulerian SPH extensions, 
 including particle relaxation to satisfy zero-order consistency, 
 kernel correction matrix to ensure first-order consistency and release the constraint associated with the normal direction of interfaces, 
 as well as dissipation limiters to enhance numerical accuracy 
 and these extensions make Eulerian SPH rigorously equivalent to FVM.
 Furthermore, 
 we implement mesh-based FVM within SPHinXsys, an open-source SPH library, 
 through developing a parser to extract necessary information from the mesh file 
 which is exported in the MESH format using the commercial software ICEM.
 Therefore, these comprehensive approaches enable a rigorous comparison between these two methods.
\end{abstract}

\end{frontmatter}
%
%
\section{Introduction}\label{referebce}
With the continuous development of high-performance computer, 
computational fluid dynamics (CFD) has been recognized as a promising approach to solve a wide range of industrial problems, 
and also to augment the understanding of many longstanding flow problems, 
from microfluidics to hydrodynamics and hypersonics \cite{afshari2018numerical,o2022eulerian}. 
While classical mesh-based CFD methods have achieved great success, 
the generation of high-quality meshes remains a major challenge in particular for complex geometries used in practical applications. 
As an alternative, 
the meshless method has attracted considerable attentions 
owing to its numerical formulation is based particles 
and independent of the topology defined by a mesh. 
As one typical example, 
smooth particle hydrodynamics (SPH), 
whose numerical approximations are based on Gaussian-like kernel function
\cite{gingold1977smoothed, lucy1977numerical}, 
has been widely applied in CFD \cite{monaghan1994simulating}, 
structural mechanics \cite{libersky1991smooth}, 
and other scientific and engineering applications \cite{zhang2022review,luo2021particle,gotoh2021entirely},
when difficulties present for the classical mesh-base methods. 

SPH can be formulated both in the Lagrangian and Eulerian frameworks 
for flow simulations. 
While the particle position is updated with velocity in Lagrangian SPH, 
it is fixed in Eulerian SPH.
The former shows obvious advantages in simulating the flows 
associated with topology changes and involving material interfaces, 
e.g. violent free-surface flow \cite{gomez2010state}, 
multi-phase flow \cite{rezavand2020weakly} and 
fluid-structure interaction (FSI) with rigid or flexible structures 
\cite{antoci2007numerical,zhang2021multi}. 
While the Lagrangian particle introduces topological flexibility,
it can lead to poor distribution, and hence, large numerical errors 
to be handled by elaborate particle regularization techniques 
\cite{quinlan2006truncation, adami2013transport, 
litvinov2015towards, vacondio2021grand}.                
With the compensation of topological flexibility, however, 
Eulerian SPH alleviates this problem as the particles are fixed  
and the initial optimal distribution is unchanged during the simulation
\cite{basa2009robustness, nasar2019eulerian}.
Therefore, it is easier to obtain more uniform, or overall less numerical errors.    
For example, 
Noutcheuwa et al. \cite{noutcheuwa2012new} and 
Lind et al. \cite{lind2015investigations, lind2016high} 
have recently demonstrated high accuracy of Eulerian SPH for incompressible flows
using a high-order smoothing kernel for interpolation. 
Another advantage of Eulerian SPH is
that the computational efficiency 
can be much higher than that of its Lagrangian counterpart
due to the fixed particles \cite{nasar2016eulerian}.

On the other hand, 
it is known in SPH community that 
the pairwise particle interaction 
using kernel-based particle formulation in SPH discretization can be 
considered as an analog of the numerical flux 
between the surface of two computational cells 
in the main-stream Eulerian mesh-based method, 
i.e. the finite volume method (FVM)
\cite{vila1999particle, litvinov2015towards, zhang2017weakly}.
Such analog has been detailed in 
Neuhauser \cite{neuhauser2014development}
so that a FVM disretization is able to reuse the Riemann solver 
developed for the arbitrary-Eulerian-Lagrangian (ALE) SPH method 
in the same software package.
However, to which level can such analog reach 
between Eulerian SPH and FVM has not been explored yet.  
Furthermore, baring with the similarities and differences,
it is still unclear whether Eulerian SPH has accountable advantage compared to FVM.

To address these issues, 
in this paper, 
we first show that the FVM formulation can be exactly implemented 
within the framework of Eulerian SPH 
developed in an open-source SPHinXsys library \cite{zhang2021sphinxsys}.
Then the performances of the two methods are rigorously 
compared with simulations of typical fully and weakly compressible flow problems.     
To that end, several extensions of Eulerian SPH have been introduced 
to improve accuracy and numerical stability.
We exploit the particle relaxation scheme \cite{zhu2021cad} 
to generate fitted-body particles along the complex geometry
and to achieve zero-order consistency for Eulerian SPH. 
We also implement a kernel gradient correction matrix to 
achieve first-order consistency \cite{bonet1999variational}.
In addition, 
we modify the dissipation limiters introduced for 
a Lagrangian SPH \cite{zhang2017weakly} 
to control the implicit dissipation 
for optimized accuracy and numerical stability. 

This paper is structured as follows: 
in Section \ref{Eulerian compressible and weakly compressible SPH} 
the Eulerian SPH formulation together with the extensions, 
and the detailed procedure for implementing FVM 
within the framework of Eulerian SPH are given. 
Rigorous comparisons on the performance between Eulerian SPH and FVM methods are give in Section \ref{Numerical results} 
and Section \ref{Summary and conclusions} presents brief concluding remarks.
All the computational codes employed in this study have been made openly accessible via the SPHinXsys repository \cite{zhang2021sphinxsys,zhang2020sphinxsys},
 which can be accessed through the following URLs: https://www.sphinxsys.org and https://github.com/Xiangyu-Hu/SPHinXsys.
%
%
\section{Methodology}\label{Eulerian compressible and weakly compressible SPH}
In this section, the governing equations for fluid dynamics are briefly summarized and 
the corresponding discretizations for Eulerian SPH are presented. 
Then, 
the Eulerian SPH extensions and the FVM implementation within the framework of Eulerian SPH are detailed.
Finally, 
the rigorous comparison between the extended Eulerian SPH and FVM is elaborated.
%
%
\subsection{Governing equations}
The Euler equation can be described by the following equation as 
\begin{equation}\label{eqs:conservation}
\frac{\partial \mathbf{U}}{\partial t}+\nabla \cdot \mathbf F(\mathbf{U})=0,
\end{equation}
where $\mathbf{U}$ and $\mathbf{F}(\mathbf{U})$ are the vector of conserved variables and the corresponding fluxes, respectively. 
In two dimensional, 
they are given by 
\begin{equation}
\label{eqs:flux}
\mathbf{U}=\left[\begin{array}{c}
\rho \\
\rho u \\
\rho v \\
E
\end{array}\right], \quad \mathbf{F}=\left[\begin{array}{c}
\rho u \\
\rho u^{2}+p \\
\rho u v \\
u(E+p)
\end{array}\right]+\left[\begin{array}{c}
\rho v \\
\rho vu \\
\rho v^{2}+p \\
v(E+p)
\end{array}\right], 
\end{equation}
respectively. 
Here, $u$ and $v$ are the components of velocity, $\rho$ the density, $p$ the pressure and $E=\frac{\rho{\mathbf{v}}^2}{2}+\rho e$ the total energy with $e$ being the internal energy.  
For compressible flow, 
we apply the equation of state (EOS)
\begin{equation}\label{getpressure}
    p=\rho(\gamma-1)e,
\end{equation}
to close the system of Eq. \eqref{eqs:conservation}. 
Here, $\gamma$ is the heat capacity ratio and the speed of sound is given by
\begin{equation}\label{sound equation}
    c^2=\frac{\gamma p}{\rho}.
\end{equation}
For incompressible flow, 
we follow the weakly compressible assumption by applying the artificial EOS 
\begin{equation}
p=c^2(\rho-\rho_0). 
\end{equation}
Here, 
$\rho_{0}$ is the reference density. 
To limit the density variation within 1\%, 
we set $c=10U_{max}$ with $U_{max}$ denoting the maximum anticipated velocity of the flow field. 
Note that the energy conservation equation in Eq. \eqref{eqs:flux} is neglected in the weakly compressible formulation.
%
%
\subsection{Standard Eulerian SPH discretization}\label{Riemann solver in discretization equations}
\begin{figure}
	\centering
	\includegraphics[width=0.9\textwidth]{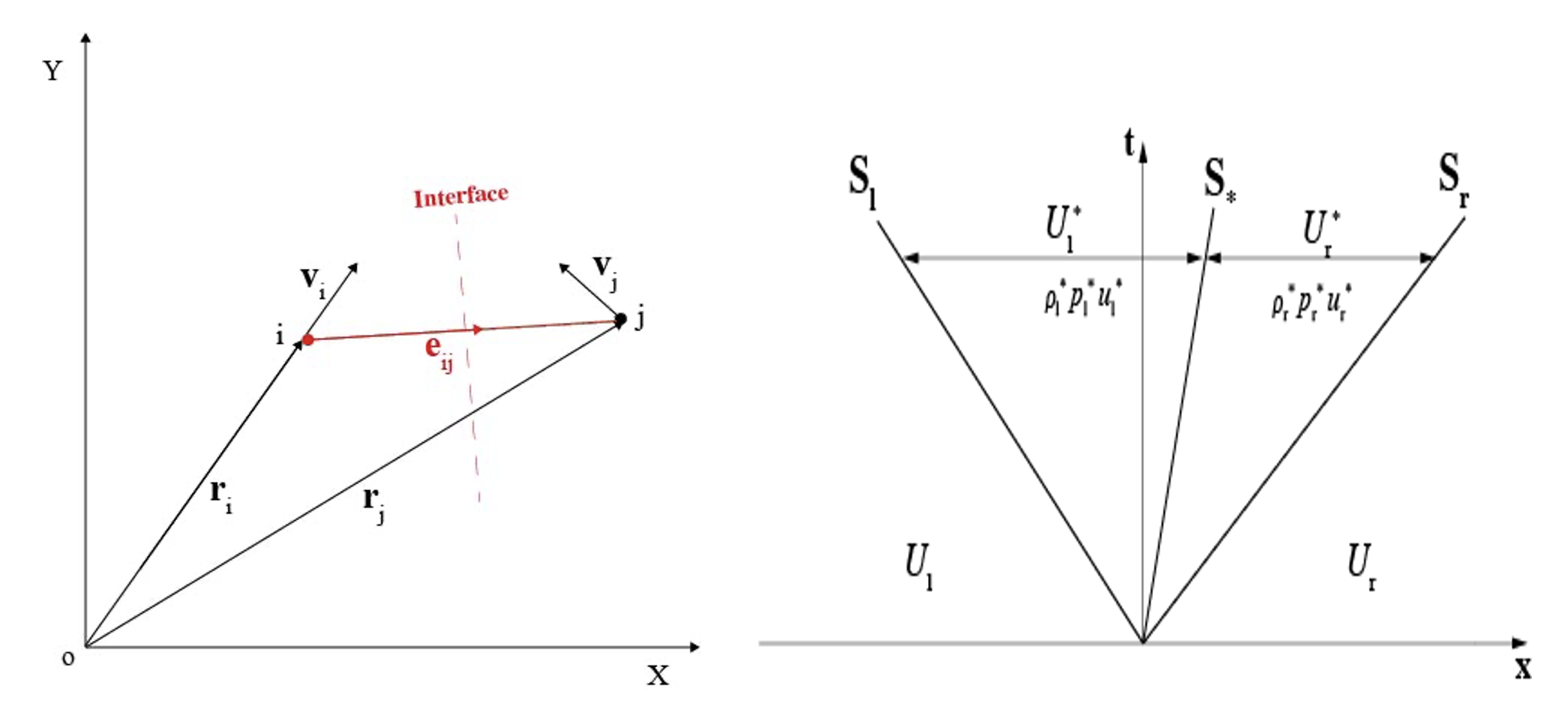}
	\caption{Construction of Riemann problem along the interacting line of particles $i$ and $j$ where 
	the interface is assumed passing the middlepoint and the normal of that is along the interaction line (left panel);
	Simplified Riemann fan with two intermediate states (right panel).}
	\label{fig:Riemann solver direction and HLLC}
\end{figure}

In the present Eulerian SPH,
the pairwise particle interaction is written in the flux form.  
Specifically,  
the mass, momentum and energy flux through the interface between a pair of particles is 
determined by solving an one-dimensional Riemann problem 
constructed along the interacting line $\mathbf{e}_{ij}$, 
as shown in Figure \ref{fig:Riemann solver direction and HLLC} (left panel). 

The left and right states of the Riemann problem are respectively 
defined as \cite{zhang2017weakly}
\begin{equation}\label{l and r states}
\left\{\begin{array}{l}
(\rho_{l},p_{l},u_{l})=(\rho_{i},p_{i},\mathbf{v}_{i}\cdot \mathbf{e}_{ij})\\
(\rho_{r},p_{r},u_{r})=(\rho_{j},p_{j},\mathbf{v}_{j}\cdot \mathbf{e}_{ij})
\end{array}, \right.
\end{equation}
assuming that the discontinuity or interface is located at the middle point $\bar{\mathbf r}_{ij}=(\mathbf{r}_{i}+\mathbf{r}_{j})/2$. 
Then, the Euler equation of Eq. (\ref{eqs:conservation}) can be discretized as \cite{vila1999particle,zhang2017weakly}
\begin{equation}\label{eqs:conservation-discretize}
\left\{\begin{array}{l}
\frac{\partial}{\partial t}\left(w_{i}\rho_{i}\right)+2 w_{i}\sum_{j} w_{j}  (\rho \mathbf{v})^{*}_{E, i j} \cdot \nabla W_{i j}=0 \\
\frac{\partial}{\partial t}\left(w_{i}\rho_{i} \mathbf{v}_{i}\right)+2 w_{i}\sum_{j} w_{j} \left[(\rho \mathbf{v} \otimes \mathbf{v})^{*}_{E, i j}+p^{*}_{E, i j}\mathbb{I}\right] \cdot \nabla W_{i j}=0 \\
\frac{\partial}{\partial t}\left(w_{i}E_{i}\right)+ 2 w_{i}\sum_{j} w_{j} \left[(E\mathbf{v})^{*}_{E, i j}+(p \mathbf{v})^{*}_{E, i j}\right] \cdot \nabla W_{i j}=0
\end{array}.\right.
\end{equation}
Here, $w$ is the volume of particle, 
$\mathbf{v}$ the velocity, $\mathbb{I}$ the identity matrix, 
$\nabla W_{i j}=\frac{\partial W_{i j}}{\partial r_{ij}}\mathbf{e}_{ij}$ 
the gradient of kernel function with the unit vector $\mathbf{e_{ij}}= -\frac{{\nabla} W_{i j}} {\left|{\nabla} W_{i j}\right|}$. 
The terms $(\rho \mathbf{v})^{*}_{E, i j}$, 
$\left[(\rho \mathbf{v} \otimes \mathbf{v})^{*}_{E, i j}+p^{*}_{E, i j}\mathbb{I}\right]$ 
and $\left[(E\mathbf{v})^{*}_{E, i j}+(p \mathbf{v})^{*}_{E, i j}\right]$, 
representing mass, momentum and energy flux, respectively, are determined from the solution of Riemann problem.

The solution of the Riemann problem results in three waves emanating from the discontinuity, 
denoted by $(\rho_l^{\ast}, u_l^{\ast},p_l^{\ast})$ and $(\rho_r^{\ast}, u_r^{\ast},p_r^{\ast})$
as shown in Figure \ref{fig:Riemann solver direction and HLLC} (right panel). 
Two waves, which can be shock or rarefaction wave, travel with the smallest wave speed $S_{l}$ or largest wave speed $S_{r}$. 
The middle wave $S_{\ast}$ is always a contact discontinuity and separates two intermediate states. 
Toro \cite{toro1994restoration,toro2019hllc} has proposed the HLLC solver based on the HLL scheme \cite{harten1983upstream}
for more accurate and robust approximation of the Riemann problem for compressible fluid flows. 
In the HLLC scheme, 
the wave speeds $S_{l}$ and $S_{r}$ estimate for the left and right regions respectively, 
are
\begin{equation}\label{sl and sr}
S_{l}=u_{l}-c_{l}, S_{r}=u_{r}+c_{r},
\end{equation}
with $c$ denoting the sound speed. 
Then, the intermediate wave speed $S_{\ast}$ is calculated as
\begin{equation}\label{eq:u*}
	S_{\ast}=\frac{\rho_{r} u_{r}\left(S_{r}-u_{r}\right)+\rho_{l} u_{l}\left(u_{l}-S_{l}\right)+p_{l}-p_{r}}{\rho_{r}\left(S_{r}-u_{r}\right)+\rho_{l}\left(u_{l}-S_{l}\right)}.
\end{equation}
Then other states in the star region can be derived as following 
\begin{equation}\label{eq:p*}
p^{*}=p_{l}+\rho_{l}\left(u_{l}-S_{l}\right)\left(u_{l}-u^{*}\right)=p_{r}+\rho_{r}\left(S_{r}-u_{r}\right)\left(u^{*}-u_{r}\right),
\end{equation}
\begin{equation}\label{eq:v*}
	\mathbf{v}_{l/r}^{*}=u^{*}\mathbf{e_{ij}}+ \left[\frac{1}{2}(\mathbf{v}_{l}+\mathbf{v}_{r})- \frac{1}{2}({u}_{l}+{u}_{r})\mathbf{e_{ij}}\right],
\end{equation}
\begin{equation}\label{eq:rho*}
\rho_{l/r}^{*}=\rho_{l/r} \frac{\left(S_{l/r}-q_{l/r}\right)}{\left(S_{l/r}-u^{*}\right)},
\end{equation}
\begin{equation}\label{eq:E*}
E_{l/r}^{*}=\frac{\left(S_{l/r}-q_{l/r}\right) E_{l/r}-p_{l/r} q_{l/r}+p^{*} u^{*}}{S_{l/r}-u^{*}},
\end{equation}
where $u^{*}=S_{\ast}$ and $q = u n_{x}+v n_{y}$ with $n_{x}$ and $n_{y}$ being components of the unit normal vector $\boldsymbol{n}$.

For weakly compressible fluid flows,
with the assumption that the intermediate states satisfy ${p}^{*}_{l}={p}^{*}_{r}={p}^{*}$ and ${u}^{*}_{l}={u}^{*}_{r}={u}^{*}$, 
a linearised Riemann solver can be derived as \cite{toro2013riemann}
\begin{equation}\label{linearised Riemann solver}
\left\{\begin{array}{l}
u^{*}=\frac{u_{l}+u_{r}}{2}+\frac{1}{2} \frac{\left(p_{l}-p_{r}\right)}{\bar{\rho} \bar{c}} \\
p^{*}=\frac{p_{l}+p_{r}}{2}+\frac{1}{2} \bar{\rho}  \bar{c} \left(u_{l}-u_{r}\right)
\end{array},\right.
\end{equation}
where $\bar{\rho}$ and $\bar{c}$ represent interface-particle averages.
With the HLLC or linearised solution to the Riemann problem, the corresponding interface flux in Eq. \eqref{eqs:conservation} can subsequently be written as 
\begin{equation}\label{choose of flux}
\boldsymbol{F}= \begin{cases}\boldsymbol{F}_{l} & \text { if } S_{l}>0 \\ \boldsymbol{F}_{l}^{*} & \text { if } S_{l} \leq 0<S_{*} \\ \boldsymbol{F}_{r}^{*} & \text { if } S_{*} \leq 0 \leq S_{r} \\ \boldsymbol{F}_{r} & \text { if } S_{r}<0\end{cases}.
\end{equation}
%
%
\subsection{Comparison between Eulerian SPH and FVM}\label{comparison_ESPH_and_FVM}
\begin{figure}
	\centering
	\begin{subfigure}[b]{0.49\textwidth}
		\centering
		\includegraphics[trim = 0cm 0cm 0cm 0cm, clip, width=0.9\textwidth]{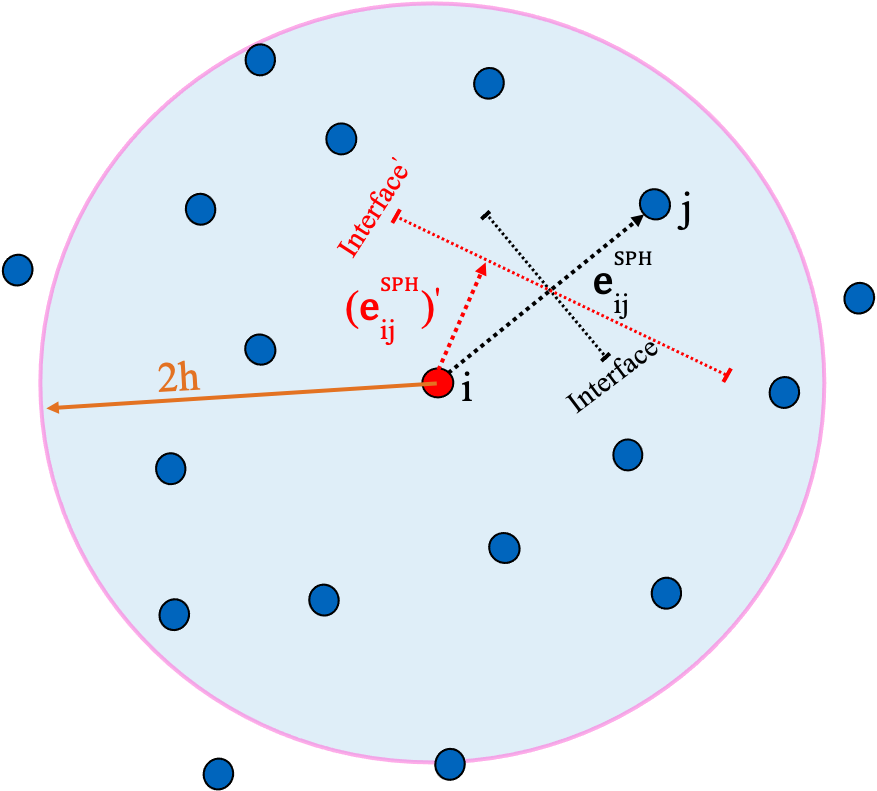}
	\end{subfigure}
	\begin{subfigure}[b]{0.49\textwidth}
		\centering
		\includegraphics[trim = 0cm 0cm 0cm 0cm, clip, width=0.9\textwidth]{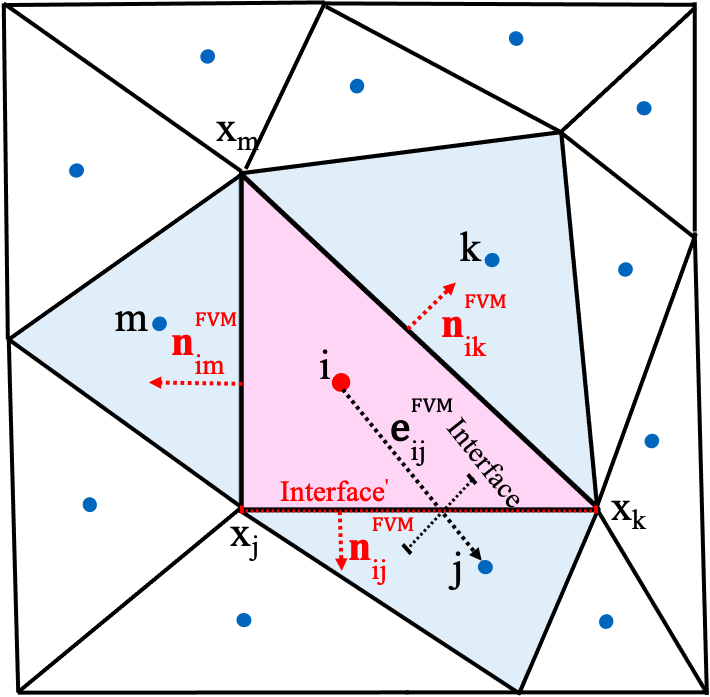}
	\end{subfigure}
	\caption{Flux Integratation: Over neighboring particles in SPH method (left panel); 
	Over neighboring cells in finite volume method (right panel).}
	\label{compasion between SPH and FVM topology}
\end{figure}

To understand the SPH formulation in Eulerian framework and its comparisons with FVM, 
we present a graphical illustration in two-dimensional between them in Figure \ref{compasion between SPH and FVM topology}. 

For the similarities between the both methods, 
they update the conserved variables by calculating the pairwise particle or cell interacting flux of all the neighbors.
In addition, by analogy with the form of the SPH discretization Eq. \eqref{eqs:conservation-discretize}, the both methods can be written uniformly as
\begin{equation}\label{sph and FVM common discretization}
	\frac{\partial}{\partial t}\left(\omega_{i} \mathbf{U}_{i}\right)+ \sum_{j} \mathbf{F}_{i j}\left(\mathbf{U}\right) \cdot \mathbf{A}_{i j}=0,
\end{equation}
Here, $\mathbf{A}_{i j}$ is a vector representing the interface area along the normal direction to the interface. 

Also, we can compare the differences between the two methods in terms of the Eq. \eqref{sph and FVM common discretization}.
The expressions of $\mathbf{A}_{ij}$ are different and denoted, respectively, as
\begin{equation}
	\label{interface area difference between sph and FVM}
	\mathbf{A}_{i j}= \begin{cases} \mathbf{A}^{ESPH}_{i j}=\left|A_{ij}\right|\mathbf{n}^{SPH}_{i j}=
		2\omega_{i} \omega_{j} {\nabla} W_{i j}= 2\omega_{i} \omega_{j}\frac{\partial W_{i j}}{\partial r_{ij}} \mathbf{e}^{SPH}_{ij}& \text { In Eulerian SPH }\\ 
		\mathbf{A}^{FVM}_{i j}=\left|A_{ij}\right|\mathbf{n}^{FVM}_{i j}={S}_{ij}\mathbf{n}^{FVM}_{i j} & \text { In FVM }\end{cases}, 
\end{equation}
where ${S}_{ij}$ is the interface size between cell $i$ and $j$, 
and the interface unit normal vector between particles $i$ and $j$ is $\mathbf{n}^{SPH}_{ij}$ in Eulerian SPH 
or cells $i$ and $j$ is $\mathbf{n}^{FVM}_{ij}$in FVM. 
In SPHinXsys library,   
the gradient of the kernel ${\nabla} W_{i j}$ is stored separately as
magnitude of gradient $\frac{\partial W_{i j}}{\partial r_{ij}}$ and displacement unit vector $\mathbf{e}^{SPH}_{ij}$ 
shown in Figure \ref{compasion between SPH and FVM topology} (left panel)
and the displacement $\mathbf{r}^{SPH}_{ij}=\left|\mathbf{r}^{SPH}_{ij}\right|\mathbf{e}^{SPH}_{ij}$ is stored as the distance $\left|\mathbf{r}^{SPH}_{ij}\right|$ 
and displacement unit vector $\mathbf{e}^{SPH}_{ij}$.
In FVM, 
we denote the center of the mesh as centroid 
and the displament of centroids $i$ and $j$ $\mathbf{r}^{FVM}_{ij}=\left|\mathbf{r}^{FVM}_{ij}\right|\mathbf{e}^{FVM}_{ij}$ is along $\mathbf{e}^{FVM}_{ij}$, 
but the fluxes through the interface in Eq. \eqref{eqs:conservation} is along $\mathbf{n}^{FVM}_{ij}$ shown in Figure \ref{compasion between SPH and FVM topology} (right panel). 
Note that interface unit normal vector $\mathbf{n}^{FVM}_{ij}$ is independent with $\mathbf{e}^{FVM}_{ij}$ in FVM. 
Therefore, a clear difference between Eulerian SPH and FVM is that 
in former the interface normal vector $\mathbf{n}^{SPH}_{ij}$ must be aligned displacement direction $\mathbf{e}^{SPH}_{ij}$ \cite{neuhauser2014development}.
A certain form of Eulerian SPH method is proposed later in section \ref{correction_matrix} to release this constraint between $\mathbf{n}^{SPH}_{ij}$ and $\mathbf{e}^{SPH}_{ij}$
and thus achieve the effect rigorously equivalent to FVM.
Besides, as is shown in Figure \ref{compasion between SPH and FVM topology} and the Eq. \eqref{interface area difference between sph and FVM}, 
the interface area $\left|\mathbf{A}_{i j}\right|$ and the way to determine neighbours between cells in FVM are given from the known grid information, 
while those in SPH method is related to the gradient and the smoothing length of the kernel.
Furthermore, each cell as a completely closed control volume naturally obeys $\sum_{j} \mathbf{A}^{FVM}_{i j}=\mathbf{0}$ in FVM, 
while the original Eulerian SPH method with approximation errors leads to $\sum_{j} \mathbf{A}^{ESPH}_{i j}\approx \mathbf{0}$, 
i.e. with consistency error \cite{neuhauser2014development}, 
which can be fixed by the particle relaxation technique mentioned later in section \ref{relaxation technique}.
%
%
\subsection{Eulerian SPH extensions} 
In this section, 
we introduce several techniques to improve numerical accuracy and stability 
and to enable extended Eulerian SPH method to be rigorously equivalent to FVM.
%
%
\subsubsection{Body-fitted particle distribution} \label{relaxation technique}
In practical applications containing complex geometry, 
the lattice particle distribution is insufficient, 
so we use particle relaxation \cite{zhu2021cad} to make the particles fit precisely on the surface of the complex geometry.

The geometry is imported before particle relaxation and the initial particles with lattice distribution are physically driven by 
a constant background pressure written as
\begin{equation}
    \mathbf{a}_{p,i}=-\frac{2}{m}_{i} \sum_{j} w_{i}w_{j} p_{0} \nabla W_{i j},
\end{equation}
where $m$ and $p_{0}$ are the mass and constant background pressure, respectively.
The particle distribution eventually reaches a steady state when $\mathbf{a}_{p,i}=\mathbf{0}$ is satisfied and then
all particles within arbitrary geometry not in the boundary after completing the relaxation satisfying
\begin{equation}
    \sum_{j} w_{j} \nabla W_{i j}=\mathbf{0}.
\end{equation}
means that the zero-order consistency is satisfied, that is, the gradient of the constant function can be correctly calculated as $\mathbf{0}$.
Therefore, Eulerian SPH method with particle relaxation remedies the approximation error to satisfy $\sum_{j} \mathbf{A}^{ESPH}_{i j}=\mathbf{0}$
in Eq. \eqref{eqs:conservation-discretize}. Besides, the particles in the boundary i.e. missing some neighboring particles do not satisfy the zero-order consistency, 
but given that the particles in the boundary are assigned to a given value and do not depend on the gradient of the kernel, 
therefore we can treat all particles in the computational domain as satisfying the zero-order consistency.
%
%
\subsubsection{Kernel correction matrix} \label{correction_matrix}
To solve the directional constraint of $\mathbf{e}_{ij}$ and $\mathbf{n}_{ij}$ in Eulerian SPH mentioned in Section \ref{comparison_ESPH_and_FVM}, 
we introduce a kernel correction matrix \cite{randles1996smoothed} that can be expressed as 
\begin{equation}
	\label{B_correction}
	\mathbf{B}_i = -\left( \sum_{j} \mathbf r_{ij} \otimes \nabla W_{i j}  w_ {j}\right)^{-1}, 
\end{equation}
which enable the particles satisfy first-order consistency, 
that is, the accurate evaluation of the gradient of a linear distributiion field.
Then the gredient of kernel can be rewritten as 
\begin{equation}
	{\nabla}^{'} W_{i j}=\frac{\mathbf{B}_i+\mathbf{B}_j}{2} \nabla W_{i j}
\end{equation}
to guarantee the momentum conservation. 

As is mentioned above, 
interface unit normal vector $\mathbf{n}^{SPH}_{ij}$ in Eulerian SPH has to be along the displacement unit direction $\mathbf{e}^{SPH}_{ij}$.
Based on this, the kernel correction matrix in Eq. \eqref{B_correction} is implemented to release the constraint 
and to adjust the normal direction of interface along $(\mathbf{e}^{SPH}_{ij})^{'}$ shown in Figure \ref{compasion between SPH and FVM topology} (left panel) 
expressed as 
\begin{equation}
	(\mathbf{e}^{SPH}_{ij})^{'}=\frac{{\nabla}^{'} W_{i j}}{\left|{\nabla}^{'} W_{i j}\right|},
\end{equation}
that is analogous to $\mathbf{n}^{FVM}_{ij}$ in FVM, thus making Eulerian SPH rigorously equivalent to FVM.
Then the modified unit normal vector of interface $(\mathbf{e}^{SPH}_{ij})^{'}$ is used to replace the original vector $\mathbf{e}^{SPH}_{ij}$ 
in Eq. \eqref{eqs:conservation-discretize}.
%
%
\subsubsection{Dissipation limiters}
Similar with the observation in Ref. \cite{zhang2017weakly}, 
directly applying the Riemann solver induces excessive numerical dissipation for the SPH method, 
dissipation limiters are introduced for the HLLC and linearised Riemann solvers to decrease the numerical dissipation, 
which are used for simulating compressible and weakly compressible fluid flows, respectively. 
In particular, 
we derive a low-dissipation HLLC Riemann solver, 
where the wave speed and pressure in the star region of Eqs. \eqref{eq:u*} and \eqref{eq:p*} are re-evaluated as 
\begin{equation}
\left\{\begin{array}{l}
u^{*}=\frac{\rho_{l}u_{l}c_{l}+\rho_{r}u_{r}c_{r}}{\rho_{l}c_{l}+\rho_{r}c_{r}}+\frac{p_{l}-p_{r}}{\rho_{l}c_{l}+\rho_{r}c_{r}}\beta^{2}_{HLLC}\\
p^{*}=\frac{p_{l}+p_{r}}{2} +\frac{1}{2}\beta_{HLLC}\left[\rho_{r}c_{r}\left(u^{*}-u_{r}\right)-\rho_{l}c_{l}\left(u_{l}-u^{*}\right)\right] 
\end{array},\right.
\end{equation}
by introducing a dissipation limiter
\begin{equation}\label{HLLC limiter}
\beta_{HLLC}=\min \left(\upeta_{HLLC} \max (\frac{u_{l}-u_{r}}{\bar{c}}, 0) , 1\right).
\end{equation}
Note that we suggest that $\upeta_{HLLC}=1$ and apply its squared value in the signal speed term for intensive dissipation control.

For the linearised Riemann solver, we also introduce a dissipation limiter to the both velocity and pressure terms 
and the linearised Riemann solver with the limiter \cite{zhang2017weakly} can then be derived as
\begin{equation}\label{acoustic Riemann solver}
\left\{\begin{array}{l}
u^{*}=\frac{u_{l}+u_{r}}{2}+\frac{1}{2} \frac{\left(p_{L}-p_{R}\right)}{\bar{\rho} \bar{c}}\beta^{2}_{linearisd} \\
p^{*}=\frac{p_{l}+p_{r}}{2}+\frac{1}{2} \beta_{linearisd} \bar{\rho}  \bar{c} \left(u_{l}-u_{r}\right)
\end{array},\right.
\end{equation}
where the dissipation limiter $\beta_{linearisd}$ \cite{zhang2017weakly} is defined as
\begin{equation}\label{acoustic limiter}
\beta_{linearisd}=\min \left(\upeta_{linearisd} \max (\frac{u_{l}-u_{r}}{\bar{c}}, 0), 1\right).
\end{equation}
Here, we suggest the parameter $\upeta_{linearisd}=15$.
%
%
\subsection{FVM within Eulerian SPH framework} 
By constructing a parser, 
we read the external mesh file format generated by the commercial software ICEM to obtain all necessary information 
to implement mesh-based FVM in SPHinXsys. 
In the mesh file, the node positions and the topological relations of all meshes can be obtained directly, 
and then other required information including the size of the interface,  
its normal unit vector and the distance between centroids can further be calculated.

In extended Eulerian SPH,
we store the kernel gradient as interface unit normal vector $(\mathbf{e}^{SPH}_{ij})^{'}$ 
and the magnitude of kernel gradient $\left|{\nabla}^{'} W_{i j}\right|$ separately mentioned in Section \ref{comparison_ESPH_and_FVM}.
Based on the relation between two methods in Eq. \eqref{interface area difference between sph and FVM}, it can be deduced that
\begin{equation}
	\label{eqs:kernel_FVM}
	\frac{\partial W_{i j}}{\partial r_{ij}}=\frac{S_{ij}}{2\omega_{i}\omega_{j}}.
\end{equation}
To implement FVM in SPH method, following the data structure of SPHinXsys,
we analogize the storage form of FVM to SPH method as also two parts including the interface unit normal vector $\mathbf{n}^{FVM}_{ij}$ 
and the magnitude of kernel gradient $\frac{S_{ij}}{2\omega_{i}\omega_{j}}$
where $\mathbf{n}^{FVM}_{ij}$ and $S_{ij}$ are calculated from the mesh information. 
Also, 
we store ${r}^{FVM}_{ij}$, the distance of centroids $i$ and $j$, in SPHinXsys storage space $r_{ij}$ and is used for solving the viscous force equation.
%
%
\subsection{Time integration}
For the time integration, 
we apply the Verlet scheme \cite{zhang2021multi} where the total energy and density are first updated to the half time step $n+\frac{1}{2}$ by
\begin{equation}
\left\{\begin{array}{l}
\mathbf{E}_{i}^{n+\frac{1}{2}}=\mathbf{E}_{i}^{n}+\frac{1}{2} \delta t\left(\frac{d \mathbf{E}_{i}}{d t}\right)^{n} \\
\mathbf{\rho}_{i}^{n+\frac{1}{2}}=\mathbf{\rho}_{i}^{n}+\frac{1}{2}\delta t\left(\frac{d \mathbf{\rho}_{i}}{d t}\right)^{n}
\end{array},\right.
\end{equation}
At this point, 
the internal energy and pressure are evaluated accordingly. 
Then, the change rate of momentum is calculated and applied to update the momentum to new time-step with
\begin{equation}
(\mathbf\rho\mathbf{v})_{i}^{n+1}=(\mathbf\rho\mathbf{v})_{i}^{n}+\delta t\left(\frac{d (\mathbf\rho\mathbf{v})_{i}}{d t}\right)^{n+\frac{1}{2}}.
\end{equation}
After that, the change rate of mass and energy are calculated. 
Finally, the energy and density for the new step are updated by
\begin{equation}
\left\{\begin{array}{l}
\mathbf{E}_{i}^{n+1}=\mathbf{E}_{i}^{n+\frac{1}{2}}+\frac{1}{2} \delta t\left(\frac{d \mathbf{E}_{i}}{d t}\right)^{n+1} \\
\mathbf{\rho}_{i}^{n+1}=\mathbf{\rho}_{i}^{n+\frac{1}{2}}+\frac{1}{2}\delta t\left(\frac{d \mathbf{\rho}_{i}}{d t}\right)^{n+1}
\end{array}.\right.
\end{equation}
In order to ensure numerical stability, the time step size is determined by 
\begin{equation}
\Delta t_{ac} =CFL\left[\frac{L}{d(c_0+U_{max})}\right],
\end{equation}
where $CFL=0.6$ and $d$ as well as $U_{max}$ represent the dimension and the maximum particle speed in the fluid field, respectively.
Here, $L$ denotes the smoothing length of kernel in Eulerian SPH or the minimum distance between mesh nodes in mesh-based FVM.
%
%
\section{Numerical results}\label{Numerical results}
In this section, 
a set of numerical examples including both compressible and weakly compressible flows 
are considered herein to investigate the accuracy and stability of extended Eulerian SPH method 
and its rigorous comparisons with mesh-based FVM. 
For all tests, 
Wendland kernel \cite{wendland1995piecewise} with a smoothing length $h=1.3d_p$, 
where $d_p$ is the initial particle spacing, 
is applied in Eulerian SPH method. 
For clarity, 
Eulerian SPH and extended Eulerian SPH, 
which means the former coupled with Eulerian SPH extensions, 
are denoted as "ESPH" and "EESPH", respectively.
Also, mesh-based FVM with the same Riemann solvers as well as the dissipation limiters is denoted as "FVM". 
Note that we visualize the mesh information by changing the VTK file in SPHinXsys
and the results after interpolation in Post-processing software called Paraview.
%
%
\subsection{Double Mach reflection of a strong shock}
In this section, 
we test a two-dimensional problem namely double Mach reflection of a strong shock 
to rigorously compare EESPH with FVM in the compressible flow. 
Following Ref. \cite{woodward1984numerical}, 
the computational domain is $(x,y) \in [0,4]\times[0,1]$ and the initial condition is given by
\begin{equation}
	(\rho,u,v,p)= \begin{cases}(1.4,0,0,1) &  y\leq1.732(x-0.1667) \\ (8,7.145,-4.125,116.8333) & \text { otherwise }\end{cases},
\end{equation}
the final time is $t = 0.2$.
\begin{figure}
	\centering
	\begin{subfigure}[b]{1.0\textwidth}
		\centering
		\includegraphics[trim = 0cm 0cm 0cm 0cm, clip, width=0.95\textwidth]{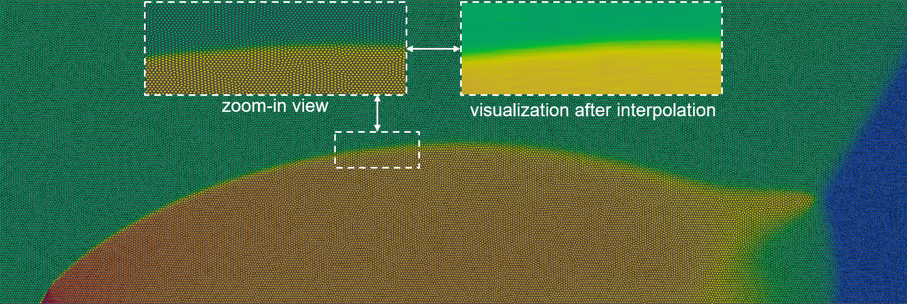}
	\end{subfigure}
	\begin{subfigure}[b]{1.0\textwidth}
		\centering
		\includegraphics[trim = 0cm 0cm 0cm 0cm, clip, width=0.95\textwidth]{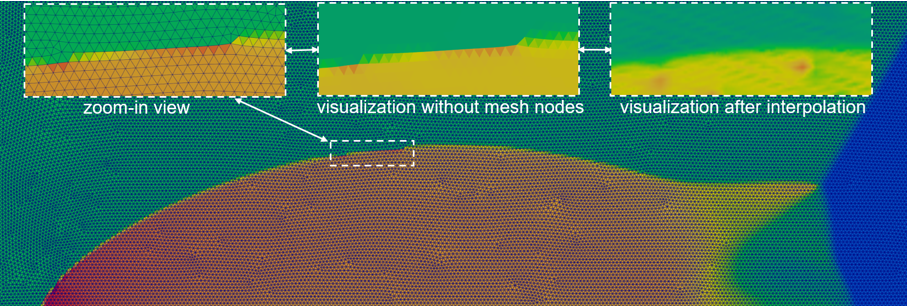}
	\end{subfigure}
	\caption{Double Mach reflection of a strong shock: 
	Density contour and its zoom-in view ranging from $1.3$ to $23.0$ obtained by EESPH (top panel)
	with the resolution $dp=1/128$, i.e. the total particles number approximately $N=6.5\times10^4$,  
	and by FVM (bottom panel) with the total elements number approximately $N=6.4 \times 10^{4}$ at the finial time $t=0.2$.
	Besides, the mesh visualization and its zoom-in view with and without mesh modes 
	as well as the visualization after interpolation in FVM are also presented.}
	\label{Density contour ranging from 0 to 25.0 obtained by ESPH and FVM.}
\end{figure}
Besides, a right moving Mach $10$ shock is initially located at $(x,y) = (0.1667,0)$ and keeps a $60$-degree angle with the $x$-axis. 
The bottom boundary is the reflective wall boundary beginning $x = 1/6$ to $x = 4$, 
the left-hand boundary is the post-shock boundary condition and zero-gradient condition is applied for the right boundary $x = 4$.
In the case, the spatial resolution is $dp=1/128$ with the total particles number approximately $N=6.5\times10^4$ in EESPH and 
the maximum element seed size is $0.012$ with the total elements number approximately $N=6.4 \times 10^{4}$ in FVM to discretize the computational domain.

In Figure \ref{Density contour ranging from 0 to 25.0 obtained by ESPH and FVM.}, 
top panel is the density contour and its zoom-in view ranging from $1.3$ to $23.0$ 
obtained by EESPH with the resolution $dp=1/128$ at the finial time $t=0.2$ 
and bottom panel presents that obtained by FVM with the total elements number approximately $N=6.4 \times 10^{4}$ 
and the mesh visualization and its zoom-in view using FVM are also given.
It can be observed that the main flow features including the Mach stem and the near-wall jet can be captured well in both methods.
In the meanwhile, 
compared with FVM, 
EESPH method has the significant advantage of obtaining a smooth density contour without any noise, 
due to the fact that in FVM the mesh has anisotropic features that lead to uneven mesh distribution, 
while EESPH method based on the isotropic kernel naturally has isotropic features.
It is important to emphasize that there are even more pronounced noises in the density contour acquired through FVM after the interpolation process in Paraview 
due to its significant gradient variation of density present in contrast to the smoother gradient observed in EESPH. 
Notably, 
the interpolation algorithm has considerably sensitivity to the variation in gradient.
%
%
\subsection{Lid-driven cavity flows with different shapes}
\begin{figure}
    \centering
    \includegraphics[width=1.0\textwidth]{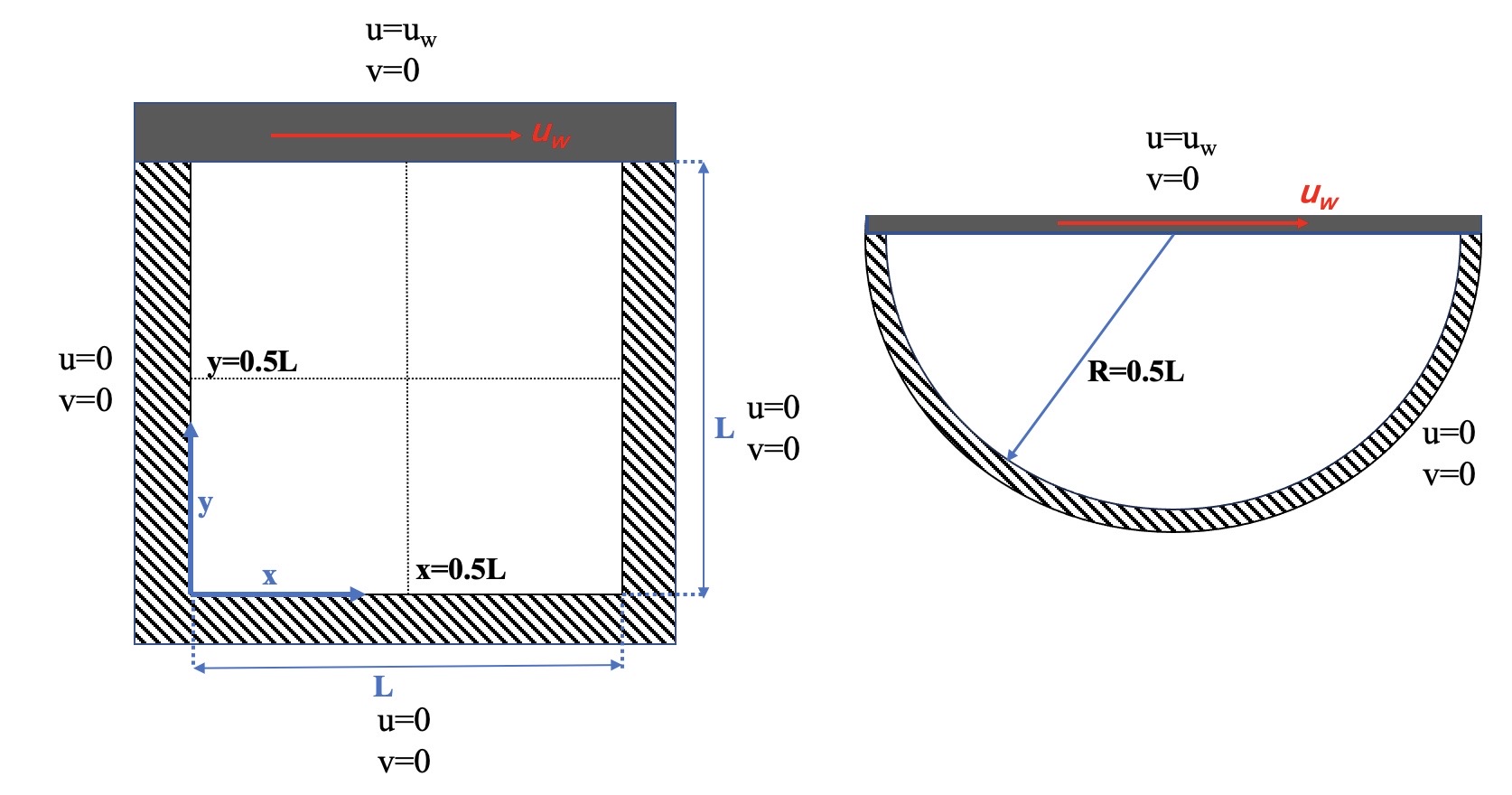}
    \caption{Lid-driven cavity flows with different shapes: Geometries and boundary conditions.}
    \label{Lid-driven cavity flow: geometries and boundary conditions}
\end{figure}

In this section, 
we consider two-dimensional lid-driven cavity flows with different shapes 
to compare EESPH with FVM further.
Firstly, a simple square cavity is applied 
and the results calculated by both methods are compared with the results from Ghia et al. \cite{ghia1982high} 
to verify its correctness. 
Besides, 
a semi-circular cavity is tested to validate the capacity to deal with complex geometry 
by comparing with the reference result from Glowinski et al. \cite{glowinski2006wall}.
The geometries and boundary conditions are shown in Figue \ref{Lid-driven cavity flow: geometries and boundary conditions}
where the upper moving wall is set as a given velocity $U_{wall}=1.0$ and other boundaries are non-slip wall conditions,
and the finial time $t=30$.
%
%
%
\subsubsection{Lid-driven square cavity problem}
For the square cavity, 
the computational domain is a square with a length of $L=1$ and the Reynolds number $Re=400$ in the case.
In EESPH,
the spatial resolutions $dp=1/33$, $1/65$ and $1/129$ are applied
with total particles numbers $N=1089$, $4225$ and $16641$, respectively, to verify the convergence study.
Correspondingly, in FVM, 
the maximum element seed sizes are set as $0.05$, $0.024$ and $0.012$ with the total elements numbers $N=1094$, $4110$ and $16096$, respectively.

\begin{figure}
    \centering
    \includegraphics[width=1.0\textwidth]{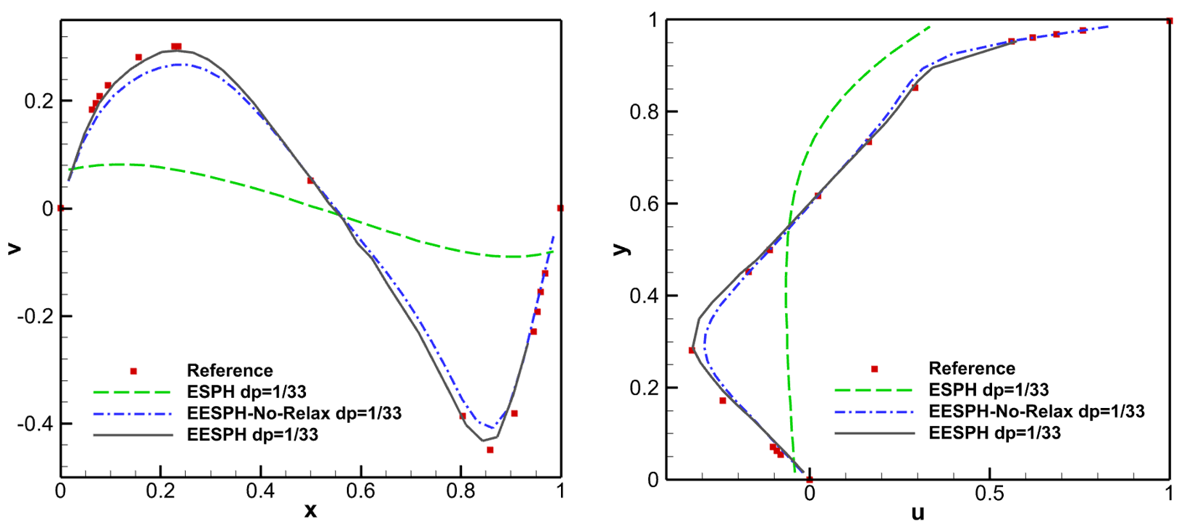}
    \caption{Lid-driven square cavity flow with $Re=400$: The horizontal velocity component $u$ along $x = 0.5L$ (left panel) and 
	the vertical velocity component $v$ along $y = 0.5 L$ (right panel) obtained by ESPH and EESPH with and without particle relaxation (denoted as EESPH-No-Relax)
	with the spatial resolutions as $dp=1/33$ and the comparison with the reference obtained by Ghia \cite{ghia1982high}.}
    \label{lid-driven-cavity-SPH}
\end{figure}
\begin{figure}
    \centering
    \includegraphics[width=1.0\textwidth]{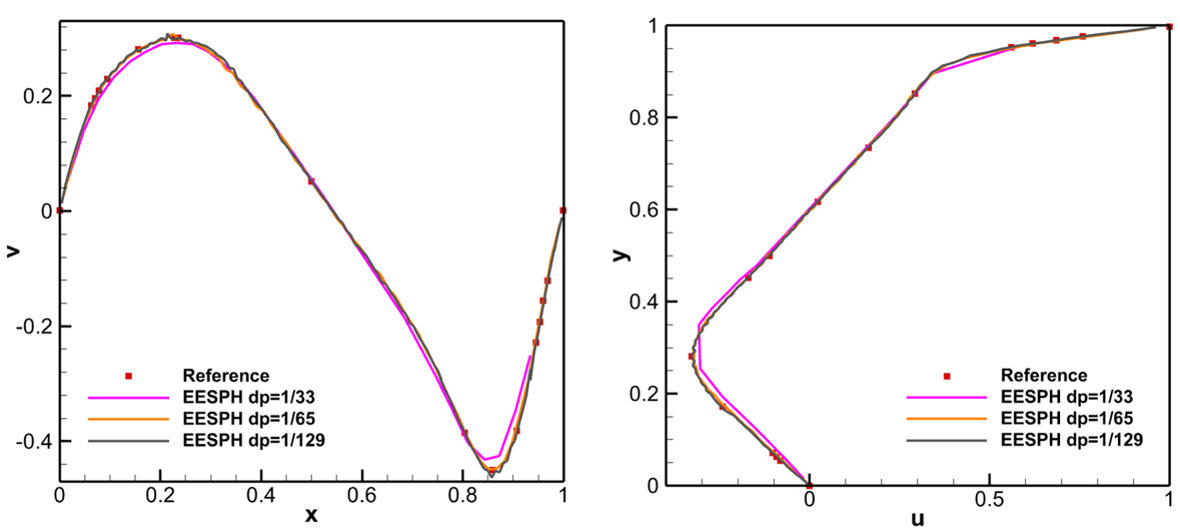}
    \caption{Lid-driven square cavity flow with $Re=400$: The horizontal velocity component $u$ along $x = 0.5L$ (left panel) and 
	the vertical velocity component $v$ along $y = 0.5 L$ (right panel) obtained by EESPH
	with the spatial resolutions as $dp=1/33$, $1/65$ and $1/129$, i.e. the total particles numbers $N=1089$, $4225$ and $16641$, 
	and the comparisons with the reference obtained by Ghia \cite{ghia1982high}.}
    \label{lid-driven-cavity-SPH-FVM}
\end{figure}
\begin{figure}
    \centering
    \includegraphics[width=1.0\textwidth]{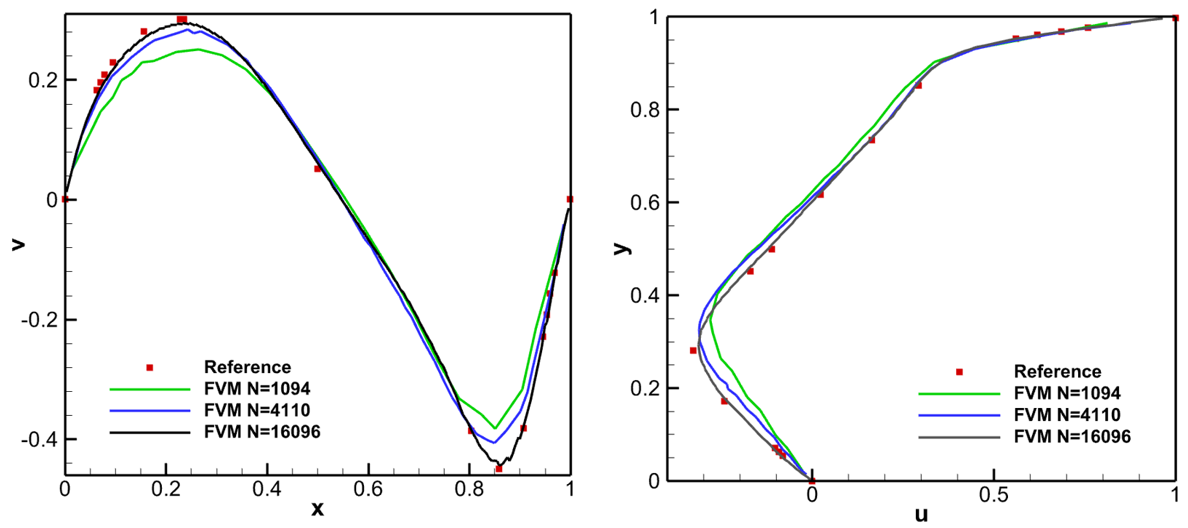}
    \caption{Lid-driven square cavity flow with $Re=400$: The horizontal velocity component $u$ along $x = 0.5L$ (left panel) and 
	the vertical velocity component $v$ along $y = 0.5 L$ (right panel) obtained by FVM with the total elements numbers $N=1094$, $4110$ and $16096$
	 and the comparison with the reference obtained by Ghia \cite{ghia1982high}.}
    \label{lid-driven-cavity-FVM}
\end{figure}
\begin{figure}
    \centering
    \includegraphics[width=1.0\textwidth]{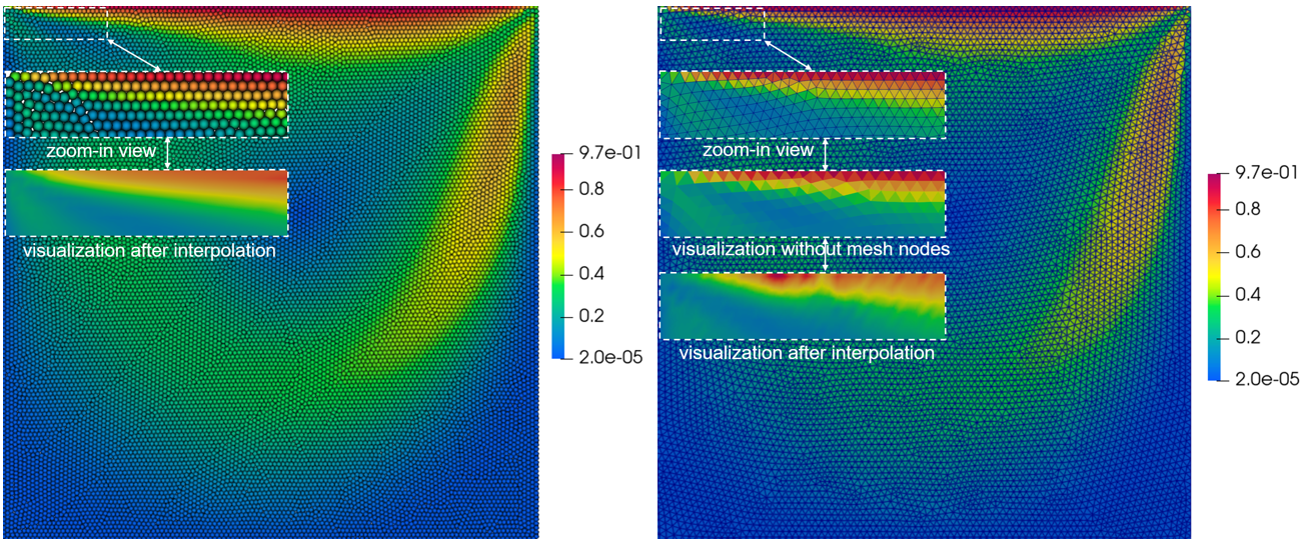}
    \caption{Lid-driven square cavity flow with $Re=400$: The velocity contour and its zoom-in view ranging from $2.0 \times 10^{-5}$ to $0.97$ 
	obtained by EESPH with the total particles number $N=16641$ (left panel) and FVM with the total elements number $N=16096$ (right panel).}
    \label{lid-driven-cavity-velocity_contout_SPH-FVM}
\end{figure}
In EESPH, 
Figure \ref{lid-driven-cavity-SPH} portrays the horizontal velocity component $u$ along $x = 0.5L$ and 
the vertical velocity component $v$ along $y = 0.5 L$ obtained by ESPH and EESPH with and without the particle relaxation (denoted as EESPH-No-Relax)
with the spatial resolution as $dp=1/33$ and the comparison with the reference obtained by Ghia \cite{ghia1982high} under the Reynolds number $Re=400$ 
in a square cavity, 
showing that Eulerian extensions can greatly improve the numerical accuracy by comparing the curves of ESPH and EESPH 
and particle relaxation can also further improve the accuracy by comparing the curves of EESPH-No-Relax and EESPH.
Besides, Figure \ref{lid-driven-cavity-SPH-FVM} presents the horizontal velocity component $u$ along $x = 0.5L$ and 
the vertical velocity component $v$ along $y = 0.5 L$ obtained by EESPH with the spatial resolutions as $dp=1/33$, $1/65$ and $1/129$ and 
the comparisons with the reference obtained by Ghia \cite{ghia1982high} in a square cavity, 
proving that the results converge rapidly with the increase of resolutions.

In FVM,
Figure \ref{lid-driven-cavity-FVM} presents the horizontal velocity component $u$ along $x = 0.5L$ and 
the vertical velocity component $v$ along $y = 0.5 L$ obtained by FVM with the total elements numbers $N=1094$, $4110$ and $16096$
and the comparison with the reference obtained by Ghia \cite{ghia1982high} with $Re=400$ in a square cavity, 
indicating that the results achieve second-order convergence as the spatial resolutions increase. 
Also, Figure \ref{lid-driven-cavity-velocity_contout_SPH-FVM} shows the velocity contour and its zoom-in view ranging from $2.0 \times 10^{-5}$ to $0.97$ 
obtained by EESPH with the total particles number $N=16641$ and FVM with the total elements number $N=16096$ under the Reynolds number $Re=400$ in a square cavity,
implying that EESPH enables obtain smooth velocity contour while results obtained by FVM are not smooth 
shown in zoom-in figure of visualization without mesh nodes and 
have more pronounced noises shown in visualization after interpolation due to the same reason explained in previous example.
In the present study, the computations are all performed on an Intel Core i7-10700 2.90 GHz 8-core desktop computer 
and the total CPU wall-clock times requred by EESPH with the total particles number $N=16641$ in whole process is $443.89s$, 
while that required by FVM with total elements number $N=16096$ is $128.49s$, 
implying that FVM is computationally much more efficient than EESPH.
%
%
\subsubsection{Lid-driven semi-circular cavity problem}
For semi-circular cavity, 
the diameter of cycle is $L=1$ and the Reynolds number is $1000$ in the case.
\begin{figure}
    \centering
    \includegraphics[width=0.92\textwidth]{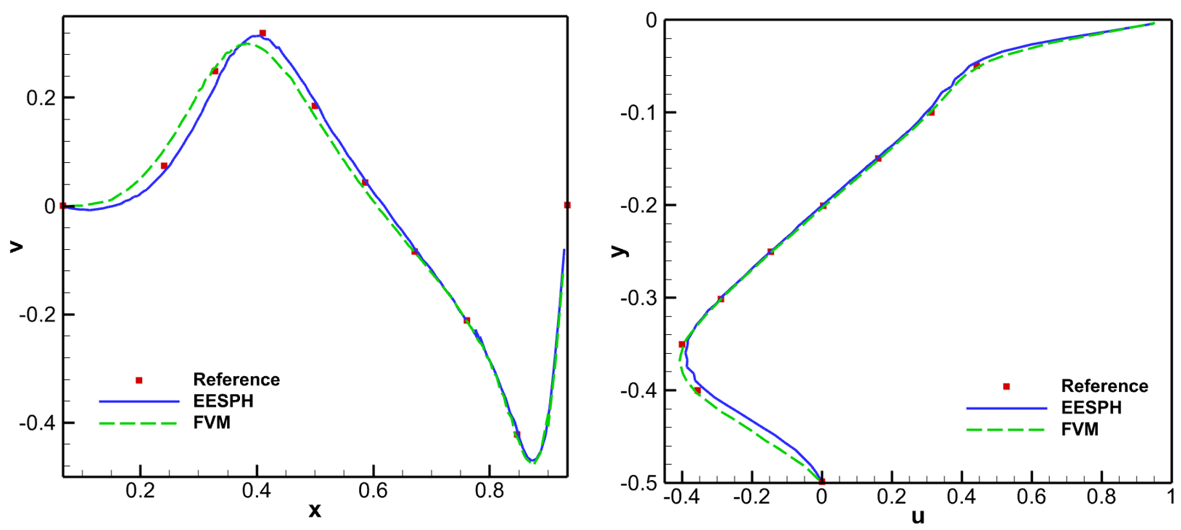}
    \caption{Lid-driven semi-circular cavity flow with $Re=1000$: The horizontal velocity component $u$ along $x = 0.5L$ (left panel) and 
	the vertical velocity component $v$ along $y = -0.25 L$ (right panel) obtained by EESPH 
	with the spatial resolutions as $dp=1/129$, i.e. the total particles number $N=6392$, 
	and FVM with the total elements number $N=6608$ and the comparison with the reference obtained by Glowinski \cite{glowinski2006wall}.}
    \label{lid-driven-semi-circular-cavity-velocity}
\end{figure}
\begin{figure}
	\centering
	\begin{subfigure}[b]{1.0\textwidth}
		\centering
		\includegraphics[trim = 0cm 0cm 0cm 0cm, clip, width=0.9\textwidth]{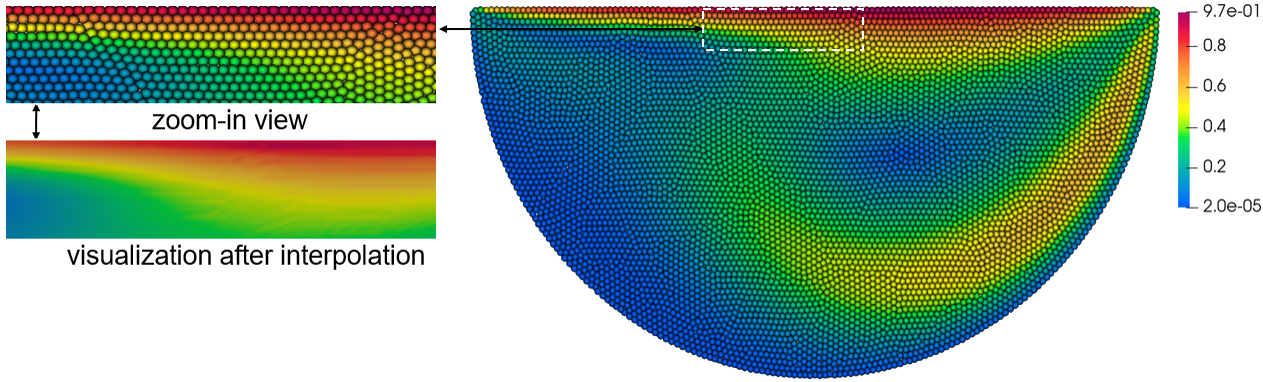}
	\end{subfigure}
	\begin{subfigure}[b]{1.0\textwidth}
		\centering
		\includegraphics[trim = 0cm 0cm 0cm 0cm, clip, width=0.9\textwidth]{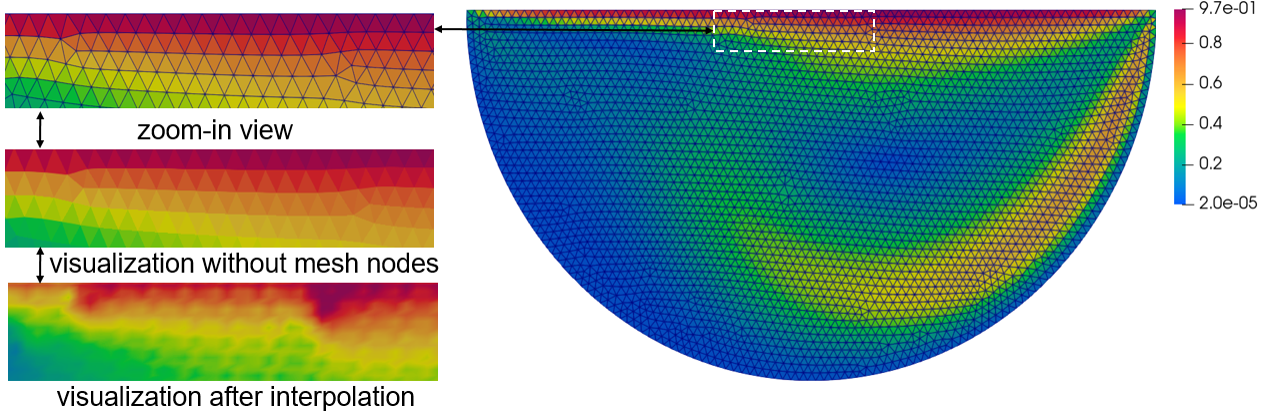}
	\end{subfigure}
	\caption{Lid-driven semi-circular cavity flow with $Re=1000$: The velocity contour and its zoom-in view ranging from $2.0 \times 10^{-5}$ to $0.97$ 
	obtained by EESPH with the spatial resolution $dp=1/129$, i.e. the total particles number $N=6392$ (top panel), 
	and FVM with the total elements number $N=6608$ (bottom panel).}
	\label{lid-driven-semi-cavity-velocity_contout_SPH-FVM}
\end{figure}
Similarly with the case above, 
we apply the resolution $dp=129$ with the total particles number $N=6392$ in EESPH 
and the total elements number $N=6608$ in FVM, respectively.
Figure \ref{lid-driven-semi-circular-cavity-velocity} shows the horizontal velocity component $u$ along $x = 0.5L$ and 
the vertical velocity component $v$ along $y = -0.25 L$ obtained by EESPH 
with the spatial resolutions as $dp=1/129$ and FVM with total number of elements $N=6608$ 
and the comparison with the reference \cite{glowinski2006wall} in a semi-circular cavity, 
proving that both methods can obtain the results which are agreement with the reference 
but the result calculated by EESPH is closer to the reference than that by FVM shown in Figure \ref{lid-driven-semi-circular-cavity-velocity} (left panel)
at the resolution $dp=1/129$.
Figure \ref{lid-driven-semi-cavity-velocity_contout_SPH-FVM} presents the velocity contour and its zoom-in view ranging from $2.0 \times 10^{-5}$ to $0.97$ 
obtained by EESPH with the spatial resolution $dp=1/129$ and FVM with the total elements number $N=6608$, 
showing that, similarly with the square cavity results, 
the result calculated by EESPH without any noise is much smoother that that by FVM with some noise 
as its anisotropic characteristics in mesh-based method.
Then we test the computational efficiency for both methods. 
The total CPU wall-clock times requred by EESPH with the particle number $N=6392$ in whole process is $168.90s$, 
while that by FVM with the total element number $N=6608$ is $50.12s$, 
showing that the computational time cost in EESPH method is much longer than that in FVM for the same physical time.
%
%
\subsection{Flow around a circular cylinder}
Furthermore, 
to rigorously compare EESPH and FVM in fluid-solid interaction, 
we investigate a case of flow around a circular cylinder as a benchmark case.
For assessing numerical results quantitatively, 
the drag and lift coefficient are defined as
\begin{equation}\label{eq:wavespeed}
C_{D}=\frac{2F_{D}}{\rho_{\infty}u_{\infty}^2 A},    C_{L}=\frac{2F_{L}}{\rho_{\infty}u_{\infty}^2 A},
\end{equation}
where $F_{D}$ and $F_{L}$ are the drag and lift forces on the cylinder respectively. 
For unsteady cases, the Strouhal number $St=fD/u_{\infty}$ with $f$ and $D$ denoting the vortex shedding frequency and the cylinder diameter, respectively. 
In the case, the computational domain is $[40D,40D]$ where the cylinder center is located at $(12.5D,20D)$
and the Reynolds numbers $Re = u_{\infty} \rho_{\infty}D/\mu$ is $100$ with the velocity $u_{\infty}=1$, density $\rho_{\infty}=1$ and the cylinder diameter $D=2$. 
Besides, all boundary conditions are the far-field boundaries and the finial time is $300$.
\begin{figure}
    \centering
    \includegraphics[width=1.0\textwidth]{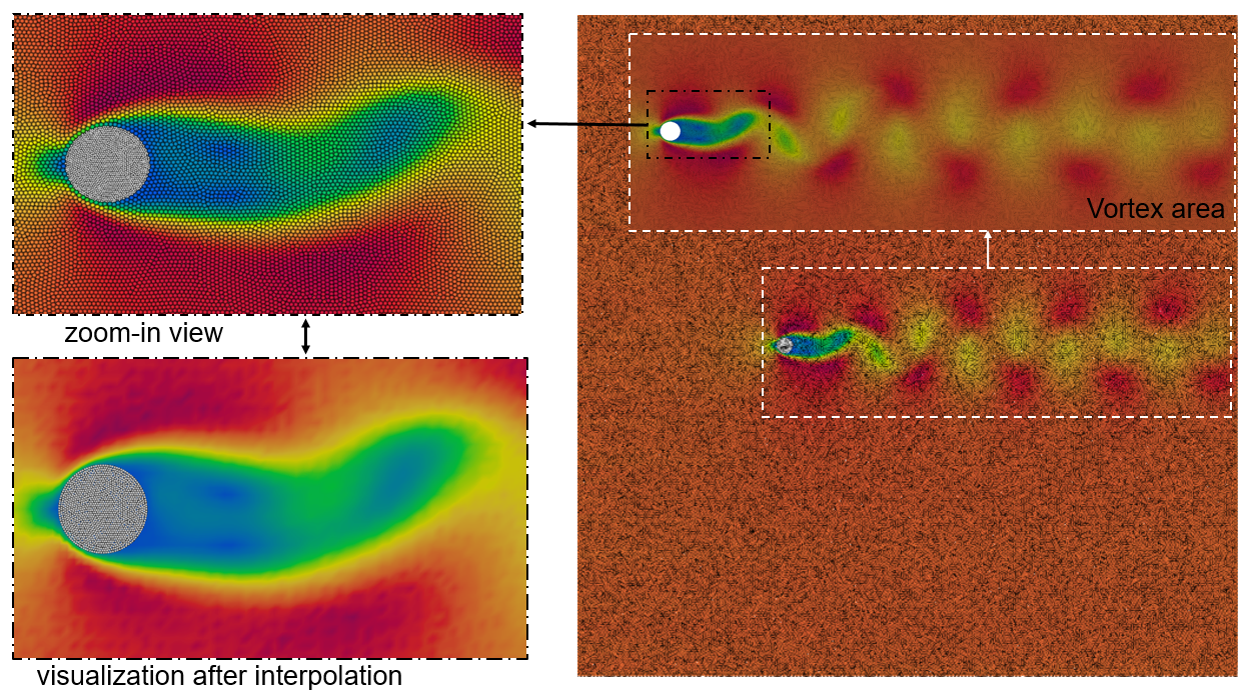}
    \caption{Flow around a circular cylinder ($Re=100$): Particle distribution and zoom-in view  as well as 
	velocity contour ranging from $0$ to $1.32$ obtained by EESPH with the spatial resolution $dp=1/10$ at the time $t=300$.}
    \label{EESPH flow around a cylinder velocity contour}
\end{figure}
\begin{figure}
    \centering
    \includegraphics[width=0.91\textwidth]{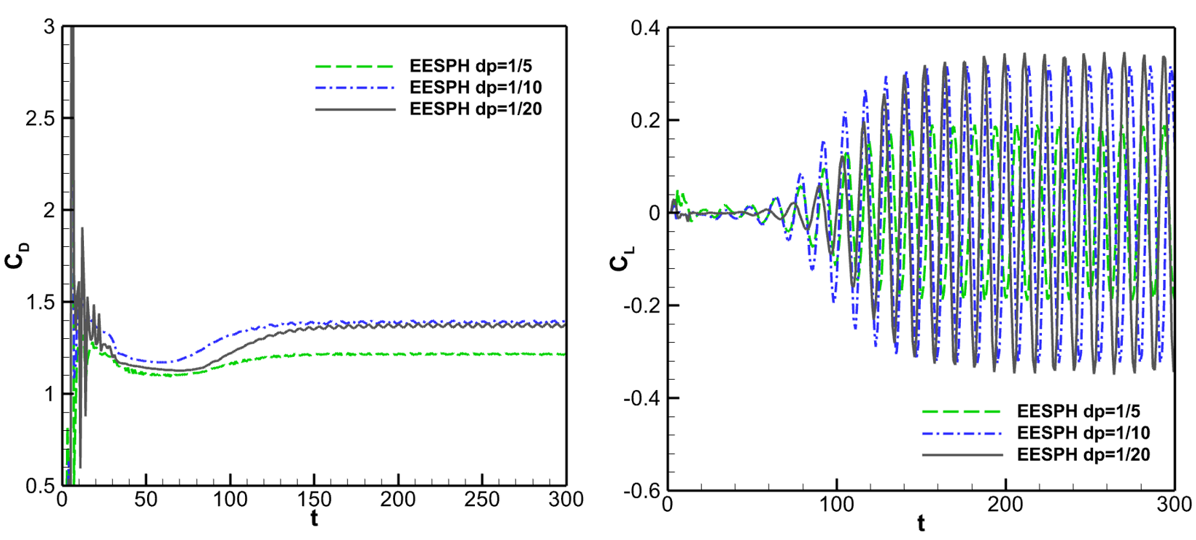}
    \caption{Flow around a circular cylinder ($Re=100$): 
	Time evolution of the drag cofficient $C_D$ (left panel) and lift cofficient $C_L$ (right panel) obtained by EESPH 
	with the spatial resolutions $dp=1/5$, $1/10$ and $1/20$.}
    \label{EESPH flow around a cylinder drag and lift coefficient}
\end{figure}
\begin{table}
    \caption{Flow around a circular cylinder ($Re=100$): Drag and lift coefficients obtainded by EESPH with the total particle number $N=2568344$ 
	and FVM with the total element number $N=2534663$ as well as other experimental and numerical results. }
    \centering
    \begin{tabular}{cccc}
    \hline
    \begin{tabular}[c]{@{}c@{}}Parameters\end{tabular} &
    \begin{tabular}[c]{@{}c@{}}$C_{D}$\end{tabular} &
     \begin{tabular}[c]{@{}c@{}}$C_{L}$\end{tabular} &
     \begin{tabular}[c]{@{}c@{}}$S_{t}$\end{tabular} \\ \hline
    White\cite{white2006viscous} & 1.46 & -  & - \\ \hline
    Chiu et al.\cite{chiu2010differentially} & 1.35 $\pm$ 0.012 & $\pm$0.303  & 0.166  \\ \hline
    Le et al.\cite{le2006immersed} & 1.37 $\pm$ 0.009 & $\pm$0.323 & 0.160  \\ \hline
    Brehm et al.\cite{brehm2015locally} & 1.32 $\pm$ 0.010 & $\pm$0.320 & 0.165  \\ \hline
	Russell et al.\cite{russell2003cartesian} & 1.38 $\pm$ 0.007 & $\pm$0.300 & 0.172  \\ \hline
	EESPH & 1.37 $\pm$ 0.011 & $\pm$0.350 & 0.178  \\ \hline
    FVM & 1.36 $\pm$ 0.006 & $\pm$0.270 & 0.164  \\ \hline
    \end{tabular}
	\label{Re=100}
\end{table}
\begin{figure}
    \centering
    \includegraphics[width=1.0\textwidth]{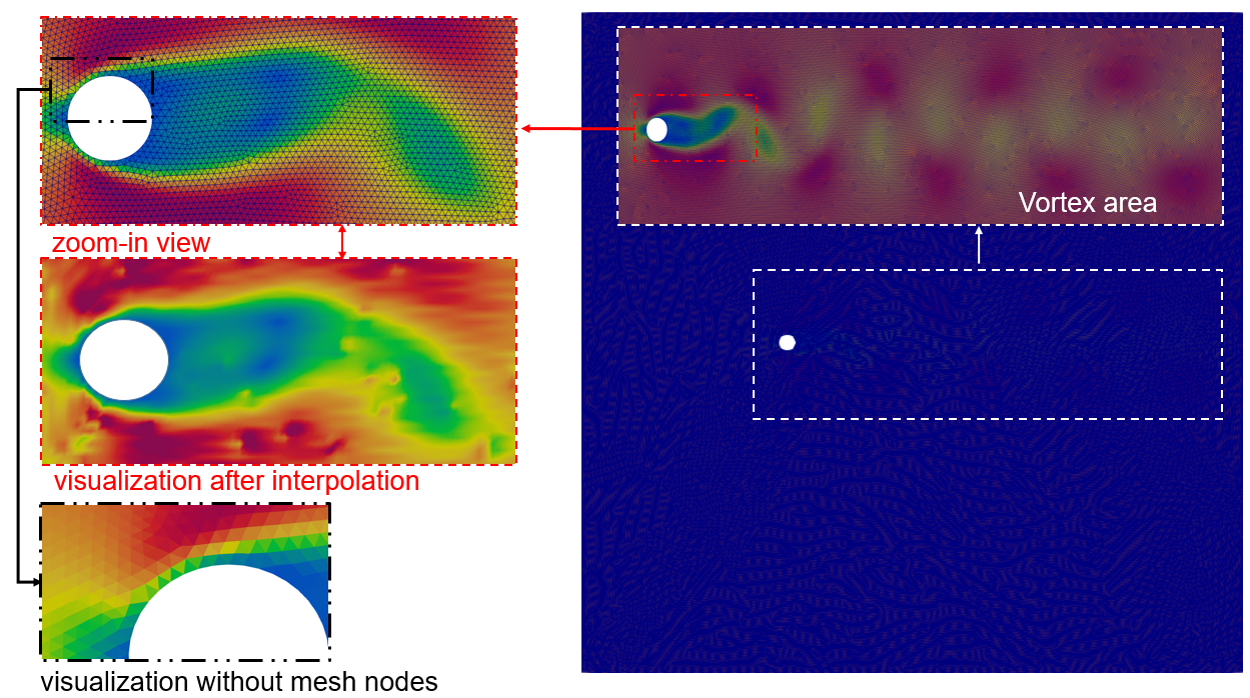}
    \caption{Flow around a circular cylinder ($Re=100$): Unstructured mesh and corresponding particle distribution
	as well as velocity contour obtained by FVM ranging from $0$ to $1.32$ with the total element number $N=639543$ at the time $t=300$.}
    \label{FVM flow around a cylinder velocity contour}
\end{figure}

In EESPH, 
the spatial resolutions $dp=1/5$, $1/10$ and $1/20$ are applied with total particle numbers $N=162328$, $644492$ and $2568344$, respectively. 
Figure \ref{EESPH flow around a cylinder velocity contour} shows the particle distribution and the zoom-in view around the cylinder as well as velocity contour 
ranging from $0$ to $1.32$ obtained by EESPH with the resolution $dp=1/10$ at the final time $t=300$, 
indicating that the velocity distribution calculated by EESPH is very smooth without any noises because of its isotropic characteristics.
Besides, Figure \ref{EESPH flow around a cylinder drag and lift coefficient} depicts the drag cofficient $C_D$ and lift cofficient $C_L$ obtained by EESPH 
with the spatial resolutions $dp=1/5$, $1/10$ and $1/20$, 
showing that the drag coefficients reach a stable mean value after a period of fluctuation at the beginning while the lift coefficient oscillates around zero.
The deviations of the drag and lift coefficients with the spatial resolutions $dp=1/10$ and $dp=1/20$ are less than 3 percent, 
and the frequencies and amplitudes of the lift coefficient are roughly the consistent, 
which means that the results are convergent.
The convergent results are listed in Table \ref{Re=100} which contains other experimental and numerical results under the Reynolds number $Re=100$, 
showing that the results obtained by EESPH agree well with other references and can be seen as correct.

Correspondingly, In FVM, 
the maximum element sizes are $0.15$, $0.076$ are applied with the total elements numbers are $639543$ and $2534663$, respectively. 
Figure \ref{FVM flow around a cylinder velocity contour} portrays the visualization of unstructured mesh, 
velocity contour and its zoom-in view around the cylinder ranging from $0$ to $1.32$ obtained by FVM 
with the total elements number $N=639543$ at the time $t=300$.
It can be seen that the velocity distribution is not smooth shown in the zoom-in figure of visualization without mesh nodes 
due to the anisotropic property of mesh-based FVM 
and much noticeable noises are observed in the visualization after interpolation due to its large gradient variation. 
Also,
the convergent result obtainded by FVM with the total elements number $N=2534663$ is also listed in Table \ref{Re=100},
implying that the drag and lift coefficients obtained by FVM are well agreement with the references. 

Furthermore, the total CPU wall-clock times requred by EESPH with the particle number $644492$ in whole process is $16059.05s$, 
while that by FVM with the total element number $N=639543$ is $7743.06s$, 
showing the much higher computational efficiency in FVM.
%
%
\section{Summary and conclusion}\label{Summary and conclusions}
In this paper, 
Eulerian SPH method is detailed and Eulerian extensions are introduced 
to make Eulerian SPH rigorously equivalent to FVM and improve the stability and accuracy of Eulerian SPH.
Also, 
mesh-based FVM is realizd in a SPH program SPHinXsys and compared rigorously with extended Eulerian SPH method.
Serval examples including fully and weakly compressible fluid flows are studied to invesitgate the different performances of extended Eulerian SPH and FVM, 
and it is concluded that both methods enable obtain correct results, 
but the former has the advantage of much smoother contours due to the isotropic property 
while the latter are is more computationally efficient due to its much less neighbours.
%
%
\bibliographystyle{elsarticle-num}
\bibliography{ref}
\end{document}